\documentclass[reprint, amsmath, amssymb, aps, prx]{revtex4-2}

\usepackage{graphicx}% Include figure files
\usepackage{dcolumn}% Align table columns on decimal point
\usepackage{bm}% bold math
\usepackage{hyperref}% add hypertext capabilities
\usepackage{physics}
\newcommand{\comment}[1]{}

\begin{document}

\preprint{APS/123-QED}

\title{Collective radiative dynamics of an ensemble of cold atoms coupled to an optical waveguide}% Force line breaks with \\

\author{Riccardo Pennetta}
\email{riccardo.pennetta@hu-berlin.de}
\author{Martin Blaha}
\author{Aisling Johnson}
\altaffiliation{Present address: Vienna Center for Quantum Science and Technology, Faculty of Physics, University of Vienna, 1090 Vienna, Austria.}
\author{Daniel Lechner}
\author{Philipp Schneeweiss}
\author{J\"urgen Volz}
\author{Arno Rauschenbeutel}
\email{arno.rauschenbeutel@hu-berlin.de}

\affiliation{%
 Department of Physics, Humboldt-Universit\"at zu Berlin, 12489 Berlin, Germany\\
}%

\date{\today}% It is always \today, today, but any date may be explicitly specified

\begin{abstract}
We experimentally and theoretically investigate collective radiative effects in an ensemble of cold atoms coupled to a single-mode optical nanofiber.
Our analysis unveils the microscopic dynamics of the system, showing that collective interactions between the atoms and a single guided photon gradually build-up along the atomic array in the direction of propagation of light.
These results are supported by time-resolved measurements of the light transmitted and reflected by the ensemble after excitation via nanofiber-guided laser pulses, whose rise and fall times are shorter than the atomic lifetime.
Superradiant decays more than one order of magnitude faster than the single-atom free-space decay rate are observed for emission in the forward-propagating guided mode, while at the same time no speed-up of the decay rate are measured in the backward direction.
In addition, position-resolved measurements of the light that is transmitted past the atoms are performed by inserting the nanofiber-coupled atomic array in a 45-m long fiber ring-resonator, which allow us to experimentally reveal the progressive growth of the collective response of the atomic ensemble.
Our results highlight the unique opportunities offered by nanophotonic cold atom systems for the experimental investigation of collective light-matter interaction.
\end{abstract}
\maketitle

%\tableofcontents

\begin{figure}[]
\includegraphics[width=0.95\linewidth]{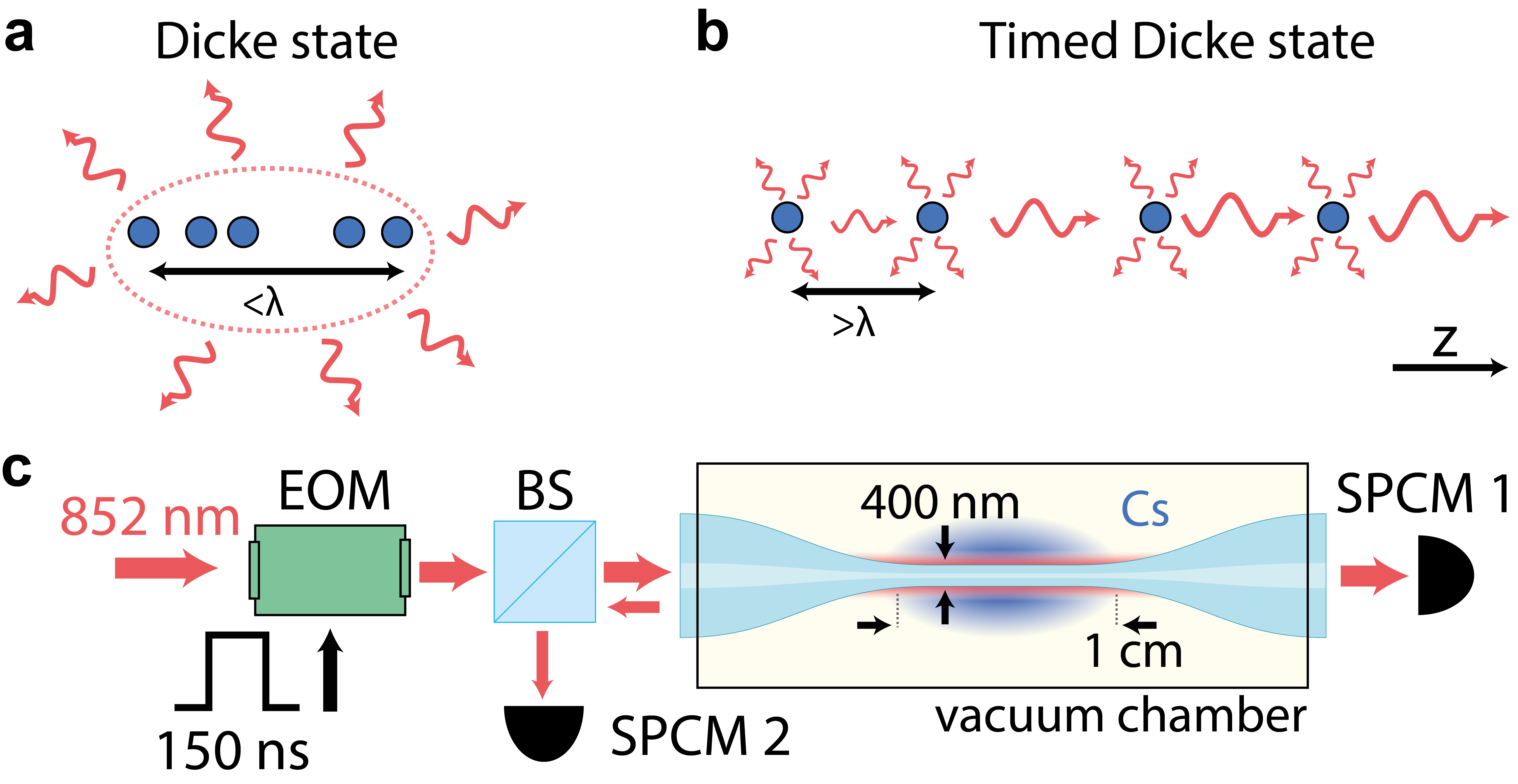}
\caption{\label{fig:ExperimentalSetup} (a-b) Sketch of a one-dimensional atomic ensemble prepared in (a) the Dicke state and (b) the timed Dicke state and corresponding emission properties. (c) Experimental setup. EOM: electro-optic amplitude modulator, BS: beam-splitter, SPCM: single-photon counting module, Cs: cesium.}
\end{figure}

\begin{figure*}[t]
\includegraphics[width=0.9\linewidth]{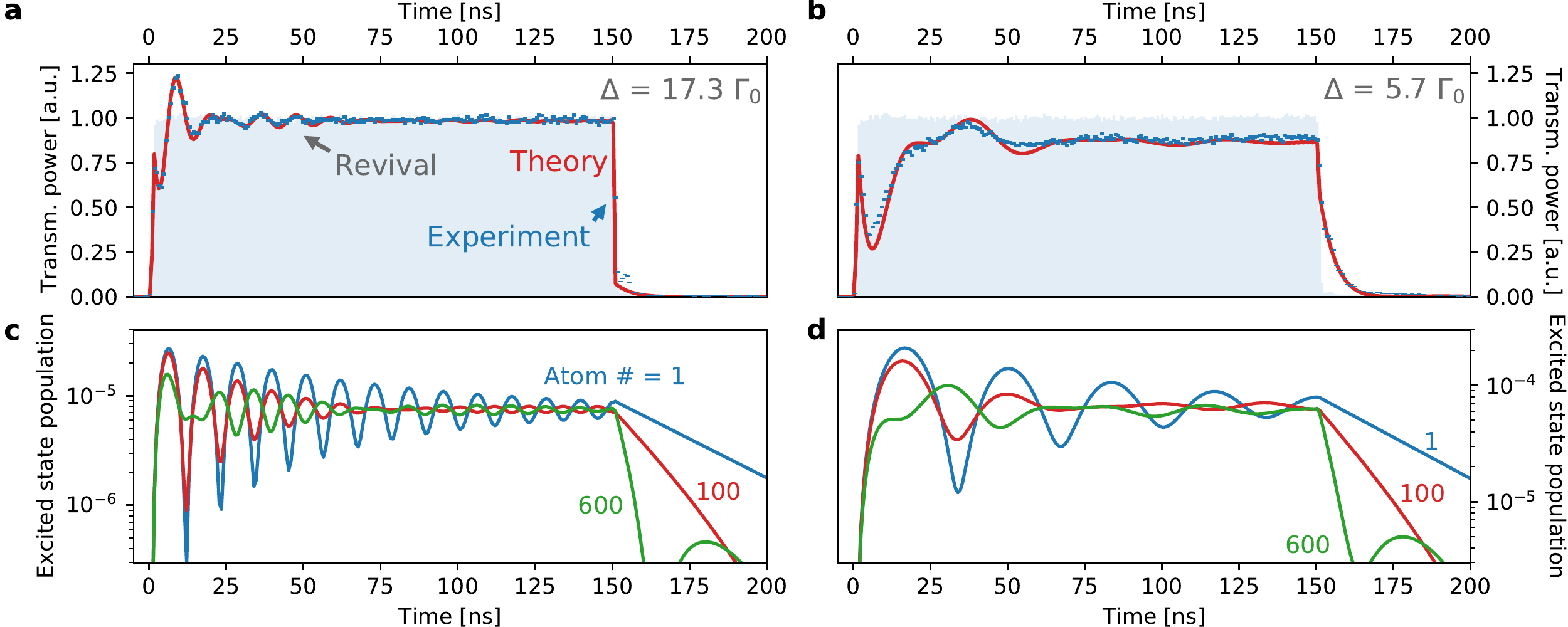}
\caption{\label{fig:TransmissionVsDetuningPlusTheory} (a,b) Measured transmission for OD = 19.3 at a detuning of (a) 17.3 $\Gamma_0$ and (b) 5.7 $\Gamma_0$. The blue dots represent the measured data, while the theoretically predicted transmission is shown as a solid red line. The blue shaded area illustrates the measured pulse shape in the absence of atoms. (c,d) Calculated probability of finding the atoms in the excited state as a function of time for the same experimental parameters of (a) and (b). The blue (respectively red and green) line indicates the excited state population probability of the 1$^\textrm{st}$ (100$^\textrm{th}$ and 600$^\textrm{th}$) atom in the ensemble.}
\end{figure*}

%\subsection{Introduction}

The study of collective effects among quantum emitters has its roots in the seminal work of Dicke \cite{Dicke1954}, which provides a clear formalism to calculate the spontaneous emission of a cloud of $N$ two-level atoms confined in a volume with dimensions small compared to the emitted wavelength, $\lambda$. In the case of a single excitation in the system, superradiant emission can be observed for an ensemble prepared in the so-called Dicke state:
%\begin{equation}
$\ket{D} =\frac{1}{\sqrt{N}} \sum_{j=1}^N \ket{j}$, 
%\label{eq:DickeState}
%\end{equation}
where the notation $\ket{j}$ indicates that the $j^\textrm{th}$ atom is excited and the others are in the ground state (see Fig. \ref{fig:ExperimentalSetup}a). 
While in the Dicke state the excitation is equally shared in the ensemble, in general terms superradiance does not require indistinguishability of the atoms, but rather of the emitted photons. In other terms, all atoms must couple to the same optical mode.

From this perspective it is not surprising that many of the phenomena first described in \cite{Dicke1954} persist even in the case of extended ensembles \cite{Scully2009}, a situation closer to many experimental configurations. For instance, super- and subradiance have been reported in cold atom clouds \cite{Araujo2016, Roof2016, Solano2017, Guerin2016, Ferioli2021, Goban2015}, Rydberg atoms \cite{ParisMandoki2017, Stiesdal2020} and ensembles of nuclei  \cite{Rohlsberger2010}. Under these circumstances, excitation through the absorption of a photon with wavevector \textbf{k} is more appropriately described by a so-called timed Dicke state  \cite{Scully2009}:
%\begin{equation}
$\ket{TD} =\frac{1}{\sqrt{N}} \sum_{j=1}^N e^{i \mathbf{k} \cdot \mathbf{r}_j} \ket{j}$, 
%\label{eq:TimedDickeState}
%\end{equation}
in which \textbf{r}$_j$ indicates the position of the $j^{th}$ atom.
Compared to the ordinary Dicke state, the introduction of the spatial phase factors breaks the symmetry of the state, causing a different dynamics for each of the atoms in the ensemble. Moreover, the system experiences an enhanced collective emission of light with wavevector \textbf{k} (i.e., in the same optical mode that excited the system) for which the different emission amplitudes interfere constructively \cite{Scully2009} (see Fig. \ref{fig:ExperimentalSetup}b).
Recent theoretical studies successfully describe the non-trivial time evolution of this state for a three-dimensional disordered atom cloud \cite{Bienaime2012, Svidzinsky2008}, however the complexity of this configuration often hinders an intuitive understanding of the dynamics of its microscopic constituents.

In this Letter we study experimentally and theoretically collective radiative effects in an ensemble of cold atoms coupled to a single-mode optical waveguide. A theoretical analysis based on a real-space quantum mechanical approach \cite{Shen2009} allows a clear microscopic (i.e., atom per atom) description of phenomena such as superradiance and collective multimode Rabi oscillations \cite{Guerin2019}. In particular, we show that the cascaded interaction among the atoms and a single guided photon causes a gradual build-up of the collective effects along the atomic ensemble in the direction of propagation of light. In contrast with the traditional Dicke description, this dynamics is independent of the inter-atomic distance (except for atoms arranged at the Bragg condition). 

We experimentally support these predictions by interfacing a cloud of laser-cooled cesium (Cs) atoms with guided photons in the evanescent field of an optical nanofiber. This configuration allows to couple thousands of atoms, whose average separation is greater than $\lambda$, to a single and well-defined guided mode and therefore represents an ideal candidate to investigate the physics of the timed Dicke state. 
We explore the temporal response of the system by exciting the atoms with boxcar shaped pulses of nanofiber-guided light, whose rise and fall times are much shorter than the atomic lifetime, and recording the power of the light that is transmitted and reflected by the ensemble.
We experimentally reveal the progressive growth of collective effects by measuring the temporal dynamics of a single optical pulse propagating multiple times through the ensemble.
This measurement, which we demonstrate to be equivalent to a single passage through atomic subsets tens of meters away from each other, allowed us to capture experimentally the position-dependent increase of the superradiant decay rate predicted by our model.

The theoretical framework used to describe light-matter interaction in our system is detailed in the supplemental material. Briefly, following the approach of Refs. \cite{Shen2009, Blaha2021}, we start with calculating the transmission amplitude in the steady-state for N atoms for a single frequency excitation:
\begin{eqnarray}
t_N( \Delta )= \prod_{j=1}^N t_i(\Delta) = \prod_{j=1}^N \left( 1 - \dfrac{\beta_j \Gamma_0}{\frac{\Gamma_0}{2} + i \Delta} \right)
\label{eq:transmission}
\end{eqnarray}
where $\Delta=\omega - \omega_a$ is the laser-atom detuning and $\beta_j$ indicates the ratio of the intrinsic spontaneous emission rate of the $j^\textrm{th}$ atom into the waveguide and the single-atom total emission rate $\Gamma_0$. In the linear regime, the time dynamics of the transmitted optical field after excitation with a pulse with scalar field amplitude $u_{in}(t)$ can be calculated as:
\begin{eqnarray}
u_{out}(t) = \mathcal{F}^{-1}[u_{in}(\Delta) \cdot t_N(\Delta)]
\label{eq:OutputPulse}
\end{eqnarray}
where  $\mathcal{F}^{-1}$ indicates the inverse Fourier transform. A similar analysis also allows to estimate the time evolution of the reflected light and the excitation amplitude for each of the atoms in the ensemble (see supplemental material). Note that Eqs. \ref{eq:transmission} and \ref{eq:OutputPulse} are independent of the position of the single atoms.

The experimental setup is sketched in Fig. \ref{fig:ExperimentalSetup}c. A cold cloud of Cs atoms from a magneto-optical trap (MOT) is prepared around a single-mode optical nanofiber (diameter $\approx$ 400 nm, waist length $\approx$ 1 cm). The atoms are probed on the Cs D2 transition ($6S_{1/2}$, $F=4 \rightarrow 6P_{3/2}, F^\prime=5$) with 150 ns long pulses of nanofiber-guided light. The pulses are generated with an electro-optic amplitude modulator (EOM) based on a Mach-Zehnder interferometer and have rise and fall times ($\approx$ 850 ps) that are short compared to the lifetime of the excited state (2$\pi$/$\Gamma_0$ = 30.4 ns, $\Gamma_0$/2$\pi$ = 5.2 MHz \cite{Steck2019}). The average emission rate of the individual atoms into the waveguide is $\beta=0.55\%$ \cite{Johnson2019}. The power of the light transmitted and reflected by the atomic ensemble is recorded using two single-photon counting modules (SPCM).

The experimental sequence starts with a preparatory phase, in which Cs atoms are loaded into the MOT for 3.0 s. Afterwards, in a cycle that is repeated a few hundred times, the MOT is released for 0.5 ms, during which 50 probe pulses are launched into the nanofiber, and then switched-on again for 200 ms to recapture and cool the Cs atoms. This sequence is repeated several hundred times allowing us to average over $\approx 10^6$ excitation  pulses. The mean power of a single pulse is much smaller than one single photon energy per atomic lifetime, placing our experiment in the linear optics (i.e., low saturation) regime.

\begin{figure}[]
\includegraphics[width=0.95\linewidth]{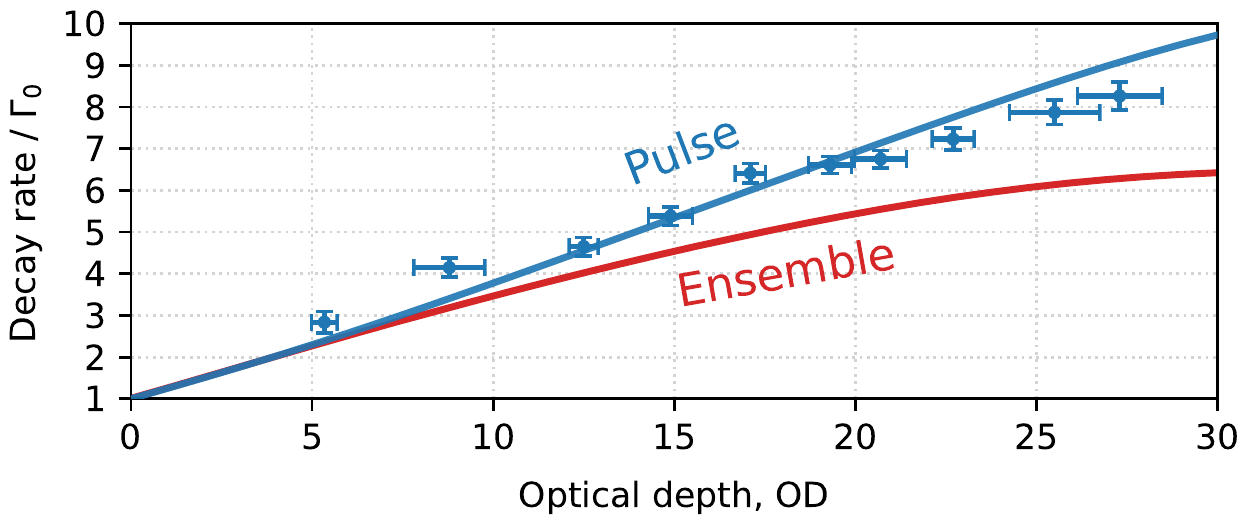}\caption{\label{fig:Superradiance} Measured (blue dots) and calculated (solid blue line) pulse decay rate as a function of the OD for a detuning of $\Delta$ = 3.8 $\Gamma_0$. The solid line depicts, the collectively enhanced decay rate of the ensemble, $\Gamma_{coll}$, i.e. the excited state population decay rate averaged over the ensemble.
}
\end{figure}

Figure \ref{fig:TransmissionVsDetuningPlusTheory} shows typical time-resolved measurements of the transmitted power for an optical depth (OD) of 19.3 (corresponding to $\approx$ 900 fully coupled atoms) and laser detuning $\Delta$ from the atomic transition of (a) 17.3 $\Gamma_0$ and (b) 5.7 $\Gamma_0$. At the leading edge of the transmitted pulse, pronounced Rabi oscillations appear, which, in contrast to the single-atom case, are not simple damped sinusoidal functions, as can be seen, e.g., from the subsequent revivals marked in Fig. \ref{fig:TransmissionVsDetuningPlusTheory}(a). These features are a clear signature of collective interaction among the atoms \cite{Guerin2019}. Theoretical predictions (without free parameters) are depicted as solid red lines and their agreement with the measurements is excellent.

To clarify the microscopic dynamics of the system, Figs. \ref{fig:TransmissionVsDetuningPlusTheory}(c-d) show the calculated time evolution of the excitation probability of the 1$^{st}$, 100$^{th}$ and 600$^{th}$ atom in the array (a more detailed plot can be found in the supplemental material).
Two features are clearly noticeable: first, while the ensemble is illuminated by the probing pulse, each atom undergoes Rabi oscillations with similar frequency (approximately equal to $\Delta$), but very different amplitude, phase and damping rate. At the beginning of the pulse, all atoms start to oscillate in phase with each other, however, the Rabi oscillations remain sinusoidal only for the first atom in the array, which behaves as if it was completely isolated from the others. The successive atoms, driven by the field that results from the interference between the probe pulse and the light emitted by all the previous atoms, eventually reverse their oscillation phase. This process may repeat itself several times for the atoms towards the end of the array. This complex dynamics is at the origin of the peculiar temporal response measured in the experiment.
Second, following the switch-off of the excitation pulse, even if the ensemble is approximately uniformly excited, the decay rate differs from atom to atom, being equal to the intrinsic rate $\Gamma_0$ for the first atom and becoming increasingly superradiant (i.e., $\Gamma > \Gamma_0$) for the subsequent ones.

To experimentally investigate the superradiant behavior in more detail, we measure the transmitted light after the switch-off of the excitation pulse as a function of the OD and infer its initial decay rate from an exponential fit, see supplemental material. Fig. \ref{fig:Superradiance} illustrates our results obtained with a laser detuning of $\Delta$ = 3.8 $\Gamma_0$ and compares them with our theoretical predictions. We observe a speed-up of the pulse decay rate of about one order of magnitude and an approximately linear dependence on the OD.
We would like to underline that the speed-up of the pulse decay rate is not a direct indicator of superradiance. The collective decay rate of the ensemble is defined as $\Gamma_{coll}(t)=-\dot{E}(t)/E(t)$, where $E$ is the total energy stored in the atoms. Therefore to calculate $\Gamma_{coll}$ we have to average the individual decay rates of the single atoms weighted by their excited state populations (see Fig. \ref{fig:TransmissionVsDetuningPlusTheory}(c,d)). The red line in Fig. \ref{fig:Superradiance} shows the calculated $\Gamma_{coll}$ at the switch-off of the excitation pulse as a function of the OD. One can see that, for the parameters used in our experiment, only for small OD the pulse decay rate is a good approximation of $\Gamma_{coll}$.

\begin{figure}[]
\includegraphics[width=0.95\linewidth]{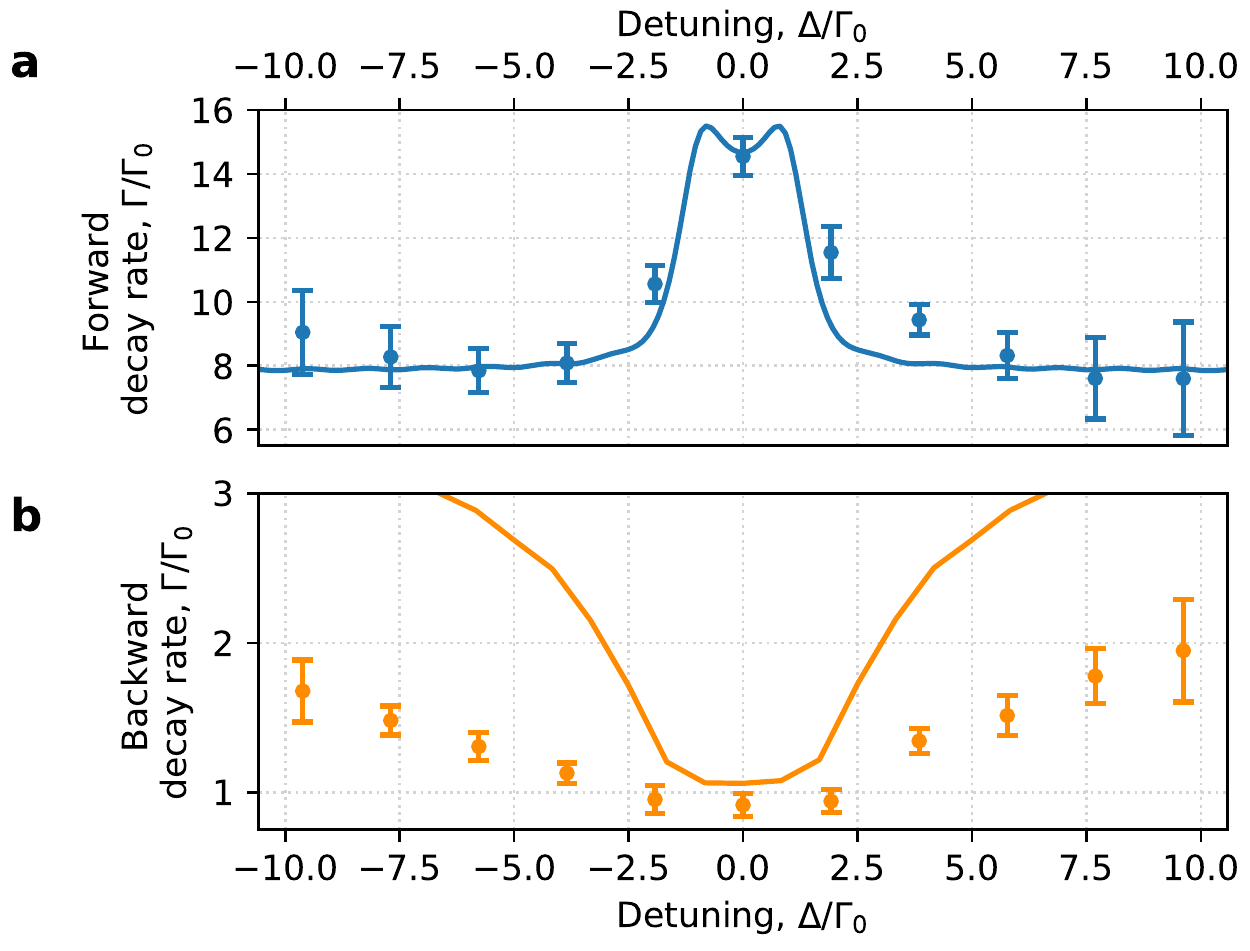}
\caption{\label{fig:BackwardForwardComparison} Measured pulse decay rate in the (a) forward and (b) backward direction for OD = 26 as a function of the laser detuning. The solid lines depict the corresponding theoretical predictions.}
\end{figure}

To further test our physical picture, we compare the pulse decay rates of the light scattered in the forward and backward direction as a function of the laser detuning (see Fig. \ref{fig:BackwardForwardComparison}). While we measure superradiant decay rates in the forward direction close to resonance, the back reflected light decays with the intrinsic rate $\Gamma_0$.
This asymmetry arises because the detected backward-propagating photons are mostly emitted by atoms at the beginning of the array.
Indeed, in a first approximation, the probability that a photon emitted from the $n^{\textrm{th}}$ atom is detected in the backward direction is $\approx rt^{\textrm{2(n-1)}}$, which decays exponentially with $n$. Here, $r$ and $t$ are the single atom reflection and transmission coefficients (see supplemental material).
Detuned excitation pulses experience a weaker light-atom coupling, resulting in a less pronounced superradiant forward decay. At the same time this allows photons reflected by atoms located deeper into the ensemble to reach the detector, which leads to an increase in the measured decay rate of the back-reflected pulse.
This behavior is qualitatively reproduced by our model (solid line in Fig. \ref{fig:BackwardForwardComparison}(b)). We attribute the discrepancies between the predicted and observed decay rate to inhomogeneous broadening of the atomic transition frequencies, e.g. due to nanofiber surface-induced detuning of the atoms \cite{Kien2007}. Its effect are negligible for the forward pulse propagation, which is dominated by the collective response of the atomic ensemble.

\begin{figure}[]
\includegraphics[width=.95\linewidth]{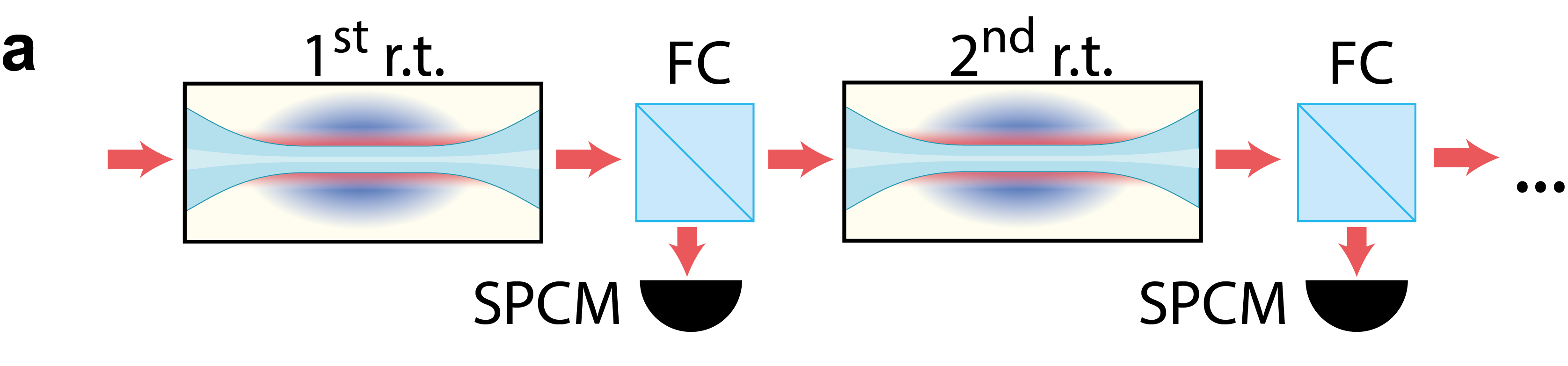}\hfill
\includegraphics[width=.97\linewidth]{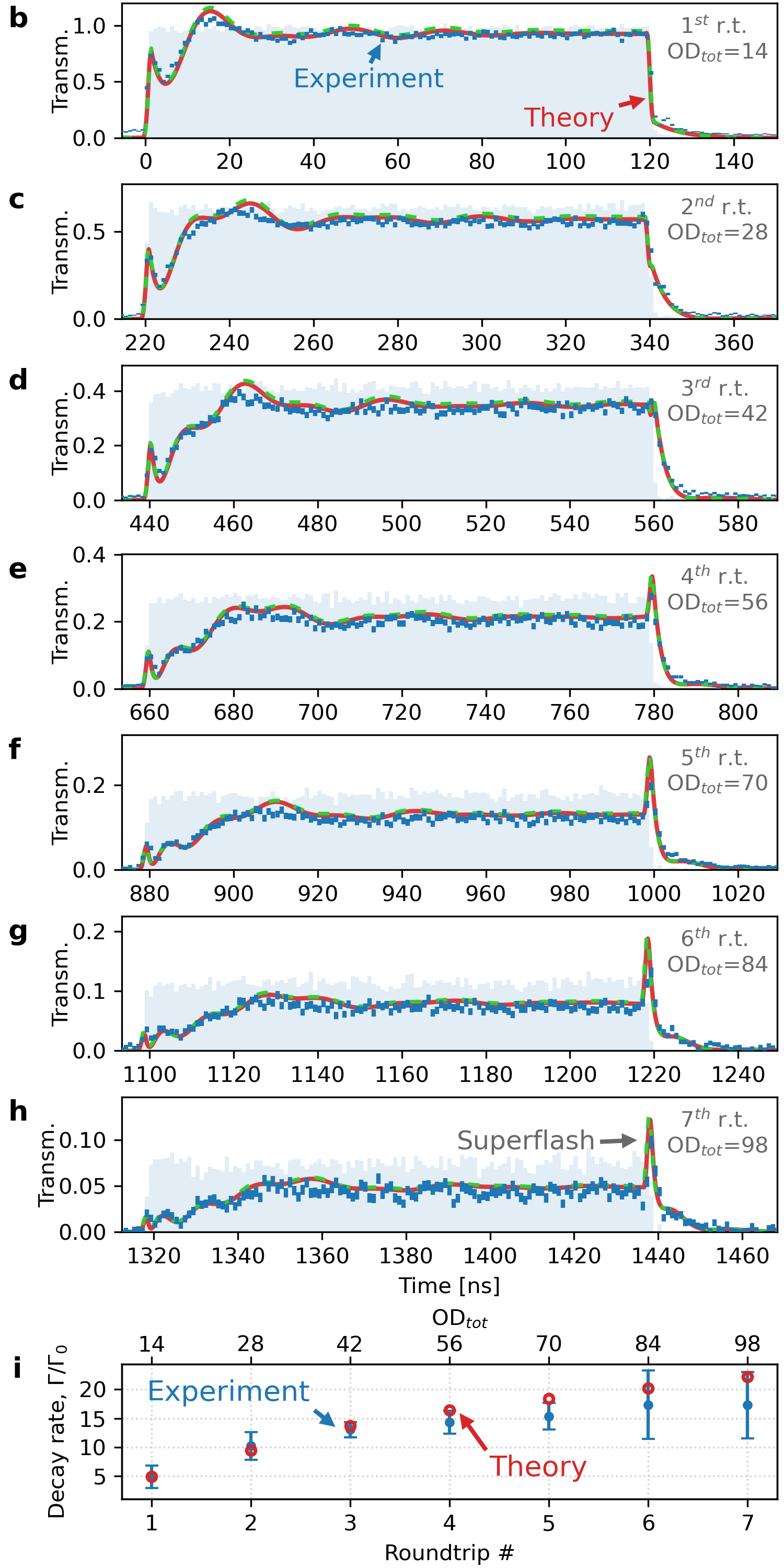}\hfill
\caption{\label{fig:Cavity} (a) Simplified schematic of the propagation of the optical pulse when the nanofiber is inserted in a $\approx$ 45-m long fiber ring resonator (FC = fiber coupler, SPCM = single photon counting module, r.t.=roundtrip). (b-h) Measured transmitted optical power of 120 ns long probe pulses for the first 7 cavity roundtrips (OD$_{sp}$=14 and detuning of $\Delta$=8.7 $\Gamma_0$). The solid red lines depict the theoretical prediction, when the presence of the optical cavity is included in the system Hamiltonian. For comparison, the green dashed lines are single-pass calculations for propagation through an optical depth OD$_{tot}$=m $\cdot$ OD$_{sp}$, where $m$ indicates the roundtrip number. (i) Measured (blue dots) and calculated (red circles) pulse decay rate as a function of the cavity roundtrip.}
\end{figure}

As mentioned, our analysis shows that collective effects in the ensemble build-up gradually along the propagation direction of light and are independent of the inter-atomic distance. This holds true even when the separation among the atoms significantly exceeds the spatial extent of the excitation pulse and the distance travelled by light in an atomic lifetime. Our nanofiber-based atom-light interface is a perfect candidate to access this regime, which, experimentally unexplored, has been subject of recent theoretical investigations \cite{Svidzinsky2008, Sinha2020}.

With this in mind, we place the nanofiber in a $\approx$ 45-m long fiber ring-resonator  \cite{Schneeweiss2016, Johnson2019}, whose cavity roundtrip time (220 ns) is much longer than the excitation pulse duration (reduced to 120 ns in the following). Under these circumstances, the ring-resonator does not provide any field enhancement but rather enables multiple passes of the probe pulse through the ensemble. At each subsequent roundtrip, part of the light is out-coupled using a fiber-coupler and detected with a SPCM. A simplified schematic of the propagation of the optical pulses is shown in Fig. \ref{fig:Cavity}(a). Thus this experimental configuration allows us to perform position-resolved measurements of the growth of collective effects in an ensemble consisting of several atomic ensembles 45 m away from each other. It is interesting to note that, since the average time of flight of the atoms through the evanescent field of the nanofiber ($\approx$ 1 $\mu$s \cite{Sague2007}) exceeds the cavity roundtrip time, the collective effects in this experiment originate from the interaction among an atomic ensemble and its time delayed counterparts. Nonetheless, since the interval between adjacent pulses is long enough for the atom cloud to completely decay into its ground state, our results after the $m^\textrm{{th}}$ roundtrip are equivalent to what could be observed with a single propagation (i.e., no optical cavity) through an ensemble with optical depth OD$_{tot}$=m$\cdot$OD$_{sp}$, where OD$_{sp}$ is the single-pass OD.

Figures \ref{fig:Cavity} (b-h) depict the measured out-coupled power for the first 7 cavity roundtrips for OD$_{sp}$ = 14 and a detuning of $\Delta$ = 8.7 $\Gamma_0$, which was obtained by averaging over $9 \cdot 10^6$ excitation pulses. The pulse switch-on dynamics is characterized by a growing complexity of the Rabi oscillations, which, roundtrip after roundtrip, increasingly deviate from the sinusoidal single-atom behaviour. In particular, for very large OD (see Fig. \ref{fig:Cavity}(e-h)) new oscillations appear, whose frequency is significantly faster than $\Delta$ and strongly depends on the OD, a regime qualitatively different from the one shown in Fig. \ref{fig:TransmissionVsDetuningPlusTheory} and discussed in Ref. \cite{Guerin2019}. At a microscopic level, this is due to the large light-matter coupling strength which causes the atoms towards the end of the array to reverse their oscillation phase before a single Rabi cycle is completed, see supplemental material.

The trailing edge of the pulses exhibits superradiant decay rates, up to approximately 17 times faster than the intrinsic decay rate $\Gamma_0$ (Fig. \ref{fig:Cavity}(i)).
Beyond OD$_{tot}$=56 we observe a change in the pulse shape and the appearance of a shoulder in the collectively emitted light field (Figs. \ref{fig:Cavity}(e-h)). The latter can be understood considering that for very large atom number, the slower decay of the atoms early in the array can re-excite the successive atoms, which then decay again at a later time.
In addition, the experiment reveals the progressive appearance of a coherent superflash of light (as referred to in Ref. \cite{Kwong2014}), whose peak intensity is larger than the one of the exciting pulse (see Fig. \ref{fig:Cavity}(e-h)).

The theoretical predictions shown in Fig. \ref{fig:Cavity} (b-h) as green dashed lines have been obtained by considering a single-pass though an ensemble with OD$_{tot}$=m$\cdot$OD$_{sp}$. We also analysed the exact experimental situation by including the optical cavity in the Hamiltonian of the system (solid red line in Fig. \ref{fig:Cavity} (b-h), see supplemental material).
The predictions for these two theoretical formulations agree for our experimental settings, meaning that our system allows us to study waveguide-mediated infinite range interactions \cite{Solano2017}.

Our results promote nanofiber-coupled atomic ensembles as a unique platform to reveal the microscopic aspects of collective effects in a one-dimensional ensemble.
We note that, while the atom-light coupling is only partially chiral in our system  \cite{Lodahl2017}, the enhancement of forward emission typical of the timed Dicke state results in properties similar to a cascaded quantum systems, in which emitters are only coupled to light which propagates in one direction \cite{Lodahl2017}.
Future research plans include extending this study beyond the single-excitation regime to explore the collective non-linear response of coupled two-level systems \cite{Cipris2021, Ferioli2021, Angerer2018}.
From this point of view, a time-resolved analysis of non-classical properties of the transmitted light, including, e.g., correlation among photons \cite{Prasad2020}, squeezing \cite{Hinney2020} and multiphoton bound states \cite{Mahmoodian2020} would certainly be of great interest.
In addition, our nanofiber ring-resonator with variable in- and out-coupling rate is an ideal candidate to investigate non-Markovian dynamics \cite{Sinha2020} as well as the physics of collective enhancement while continuously transitioning from the regime of waveguide quantum electrodynamics to cavity quantum electrodynamics.

%%%%%%%%%%%%%%%%%END OF MAIN TEXT%%%%%%%%%%%%

\begin{acknowledgments}
We acknowledge financial support by the Alexander von Humboldt Foundation in the framework of an
Alexander von Humboldt Professorship endowed by the Federal Ministry of Education and Research and by the Austrian Science Fund (NanoFiRe grant project No. P31115).
\end{acknowledgments}

\bibliography{References}% Produces the bibliography via BibTeX.

\newpage

\section*{Appendixes}

\appendix

\section{Theoretical description}

\begin{figure}[b]
\includegraphics[width=1\linewidth]{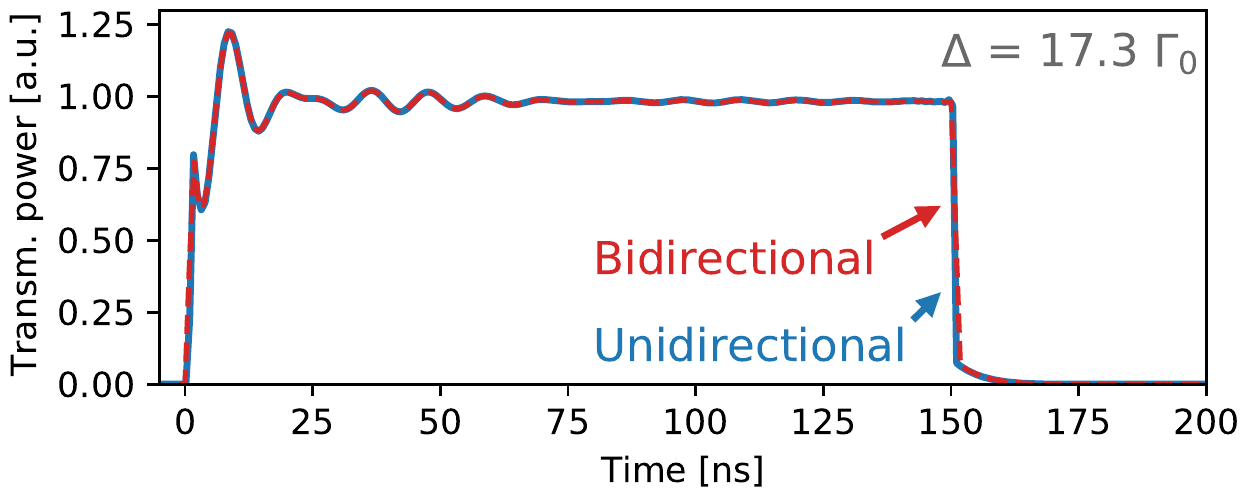}\
\caption{\label{fig:ComparisonUnidirectionalBidirectional} Comparison between shapes of the transmitted pulse predicted using the unidirectional (blue solid line) and bidirectional models (red dashed line) for  OD = 19.3 and $\Delta$=17.3 $\Gamma_0$ (same parameters as in Fig. 2a of the main manuscript).}
\end{figure}

\begin{figure*}[]
\includegraphics[width=1\linewidth]{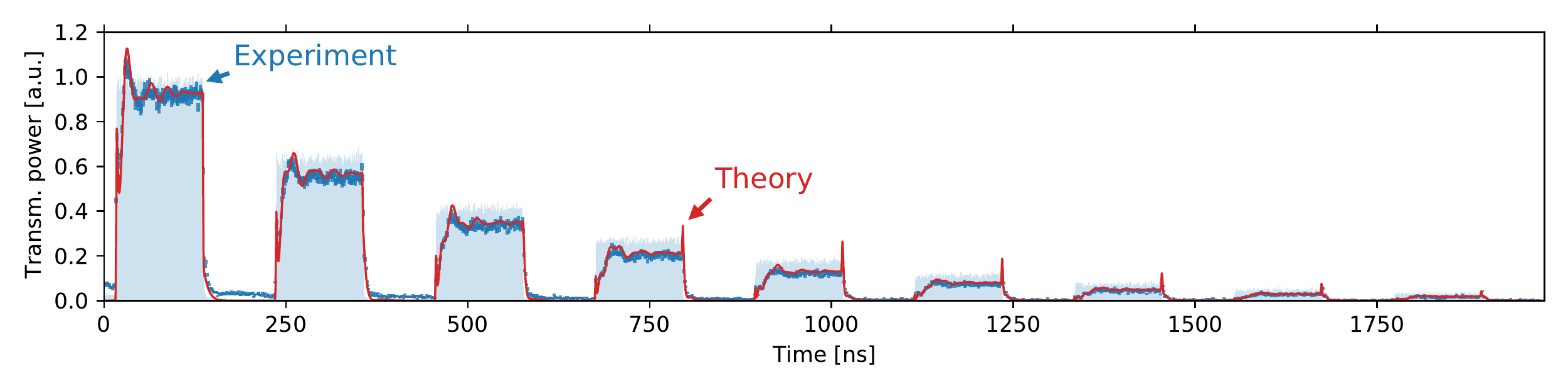}\
\caption{\label{fig:CavityComplete} Complete time trace of the cavity ringdown measurement shown in Fig. 5(b-h) of the main text (parameters: OD$_{sp}$=14 and $\Delta = 8.7 \, \Gamma_0$). The blue dots are experimental data, while the solid red line illustrates the theoretical prediction, when the presence of the optical cavity is included in the system Hamiltonian. The blue shaded area illustrates the measured cavity ringdown in the absence of atoms.}
\end{figure*}

\begin{figure}[]
	\includegraphics[width=1\linewidth]{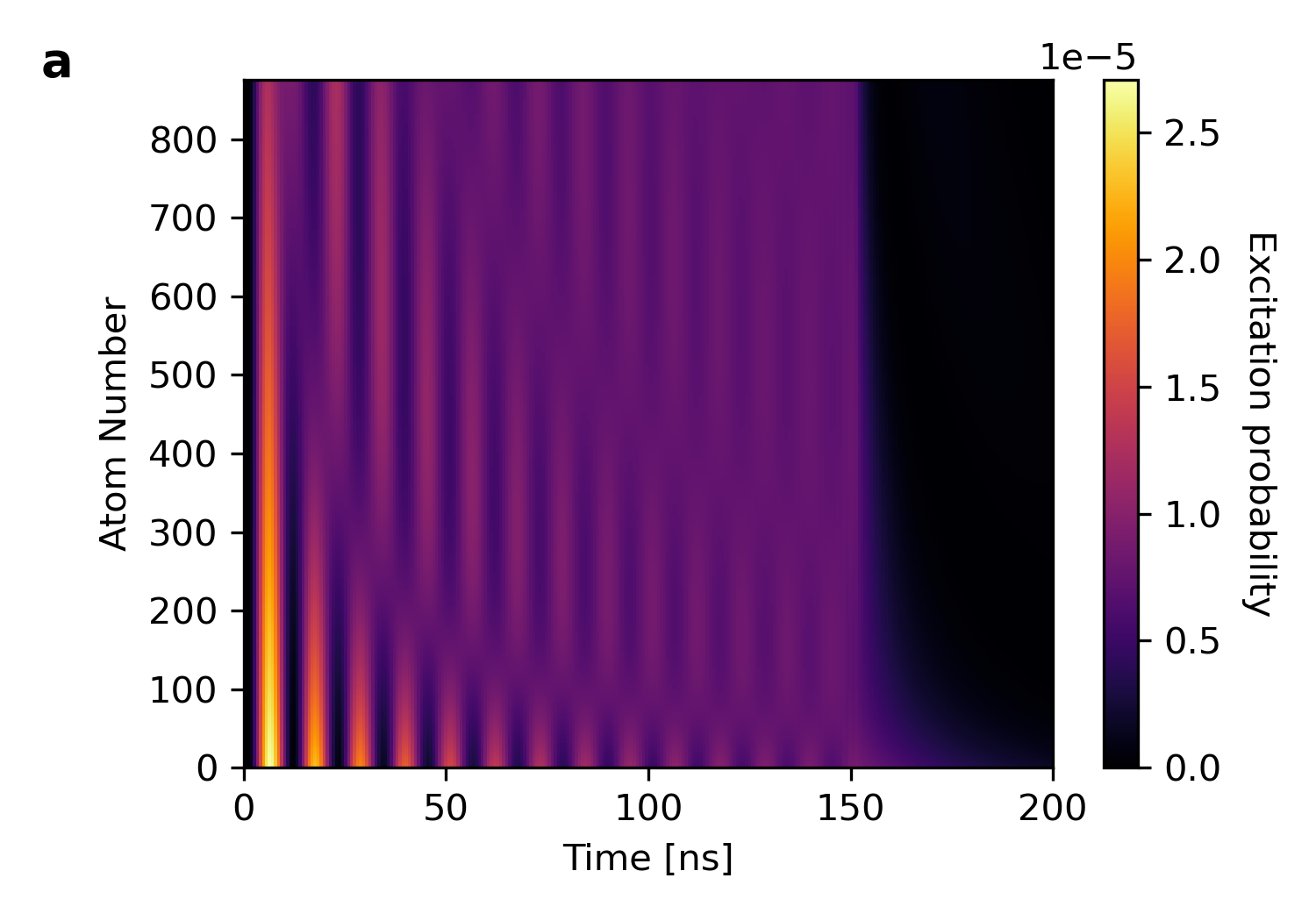}
	\includegraphics[width=1\linewidth]{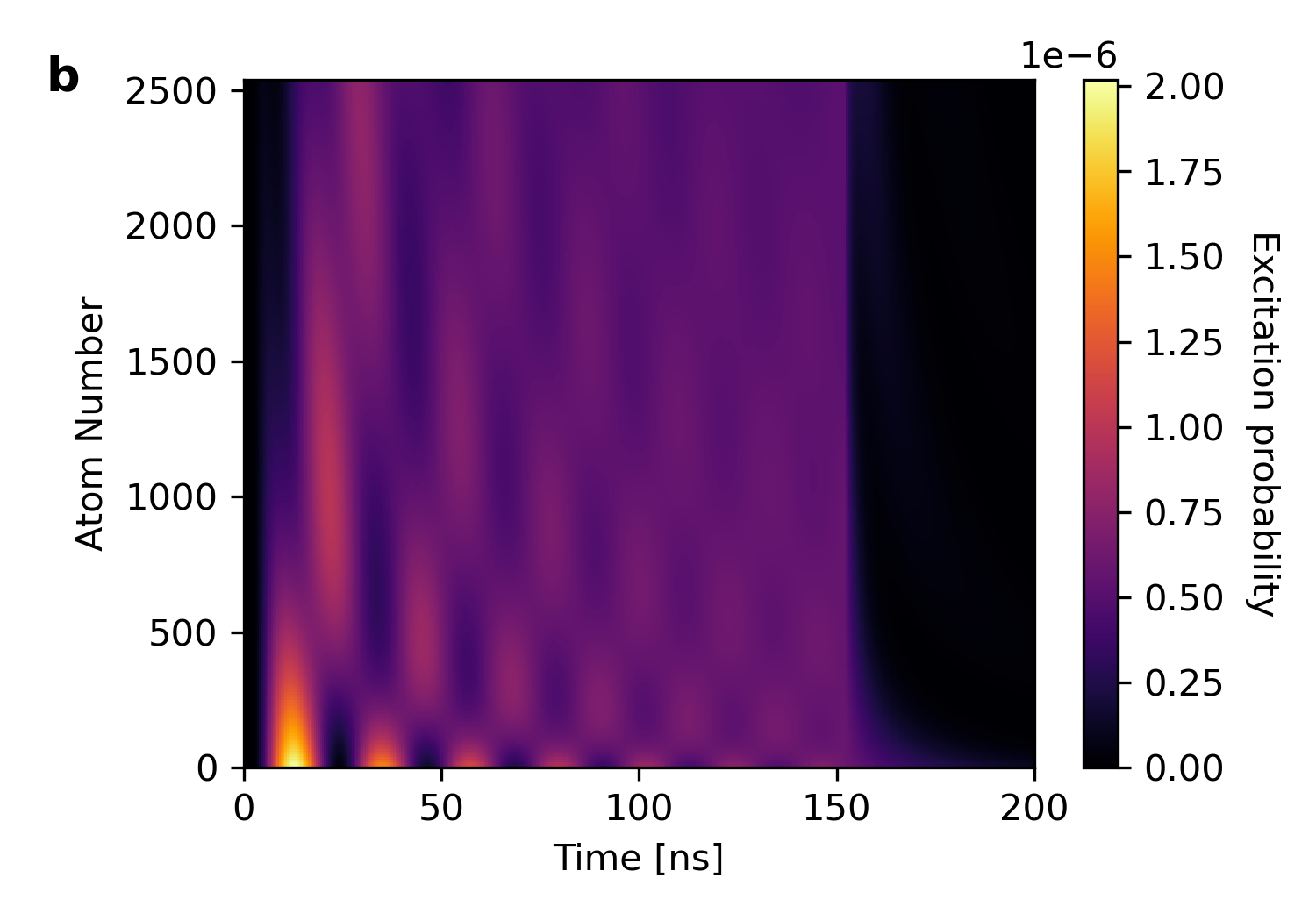}
	\caption{\label{fig:TransmissionVsDetuning_ColormapSingleAtoms} (a) Excitation probability as a function of time (colormap) for each atom at OD = 19.3 and $\Delta$ = 17.3 $\Gamma_0$ (same parameters as in Fig. 2(a) of the main manuscript). (b) Same calculation but for OD$_{tot}$ = 58 and $\Delta$ = 8.7 $\Gamma_0$ (same parameters as in Fig. 5(e) of the main manuscript). The excitation probabilities are calculated using the actual pulse energies in the experiment (2$\hbar \omega$ in (a) and 0.04 $\hbar \omega$ in (b)).}
\end{figure}

\begin{figure*}[]
	\includegraphics[width=1\linewidth]{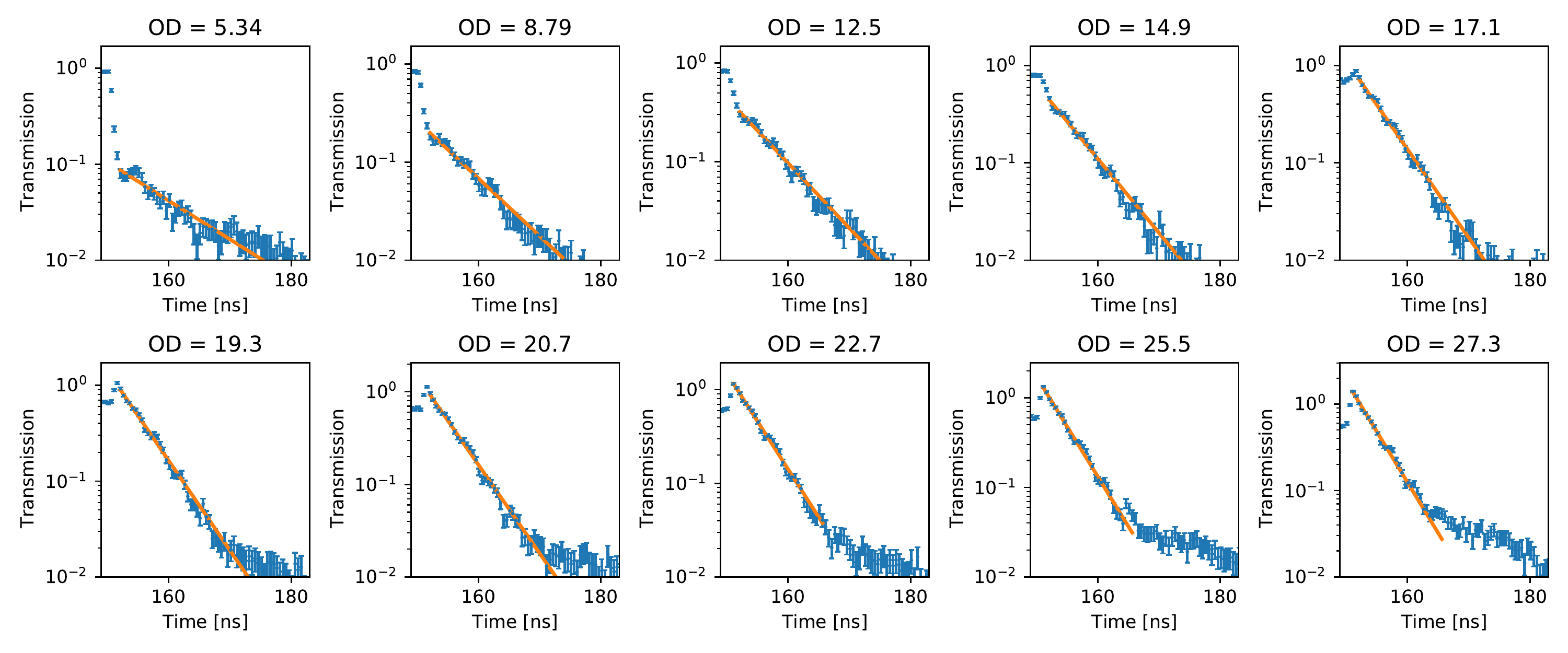}\
	\caption{\label{fig:Superradiance_AllTheFits} Data used to estimate the pulse decay rate as a function of the OD of Fig. 3 of the main manuscript. The blue dots indicate the measured transmitted power, while the orange solid lines are exponential fits. Superimposed on the pulse decay are small-amplitude oscillations that we attribute to quantum beats between the $6P_{3/2}$, $F^\prime = 4$ and $F^\prime = 5$ excited states.}
\end{figure*}

\subsection{Unidirectional model}

To describe light-matter interaction in our system, we follow the approach of \cite{Shen2009, Blaha2021} and we start by writing a real-space non-Hermitian Hamiltonian for $N$ two-level atoms coupled to a single-mode waveguide. In order to decrease the complexity of the problem, here we consider a uni-directional propagation model. Due to the collective enhancement of forward emission, this is indeed sufficient to capture most of the features observed in the experiment. In addition, here we assume that all $N$ atoms have the same coupling strength to the waveguide mode. Although this neglects fluctuations in single experimental realizations, it describes well results obtained after numerous averages. These two assumptions are dropped in the next section, where an exact solution to the problem is discussed. Under these conditions, the Hamiltonian can be written as \cite{Shen2009, Johnson2019}:
\begin{eqnarray}
\dfrac{H}{\hbar}= \int_{-\infty}^{\infty} \bigg[ a_x^{\dagger} (-i v_g \dfrac{\partial}{\partial x}) a_x +
\sum_{n=1}^N \delta(x-x_n) \Omega_a  \sigma_n^+\sigma_n^- +
\nonumber\\*
\sum_{n=1}^N \delta(x-x_n)V(\sigma_n^+a_x+\sigma_n^-a_x^{\dagger})
\bigg] dx ~,
\qquad \qquad 
\label{eq:Hamiltonian}
\end{eqnarray}
where $a_x^{\dagger}$ ($a_x$) creates (annihilates) a photon at position $x$, $\sigma_n^+$ ($\sigma_n^-$) is the raising (lowering) operator for the $n^{\textrm{th}}$ atom at position $x_n$ and $v_g$ is the group velocity of the waveguide mode. Moreover, $\Omega_a = (\omega_a-i(1-\beta) \frac{\Gamma_0}{2})$ and $V = \sqrt{v_g \beta \Gamma_0}$, where $\omega_a$ is the atomic resonance frequency and $\beta$ denotes the ratio of the spontaneous emission rate of the individual atoms into the waveguide and the total single-atom emission rate, $\Gamma_0$.

In the limit of a single excitation that is shared between the waveguide mode and the atomic ensemble, the generic state of the coupled system can be written as: $\ket{\psi} = \int_{-\infty}^{\infty} \left[ \varphi(x)  a_x^{\dagger} + \sum_{n=1}^N \delta(x-x_n) \phi_n \sigma_n^+ dx \right] \ket{0}$, where $\varphi(x)$ and $\phi_n$ are the probability amplitudes of finding a photon at position $x$ and the $n^{\textrm{th}}$ atom in the excited state, respectively. Solving the time-independent Schr\"odinger equation allows us to calculate the steady-state amplitude transmission through the ensemble for a single frequency excitation:
\begin{eqnarray}
t_N( \Delta )= t( \Delta )^N = \left( 1 - \dfrac{\beta \Gamma_0}{\frac{\Gamma_0}{2} + i (\Delta)} \right) ^N ~,
\label{eq:Transmission}
\end{eqnarray}
where $\Delta = \omega - \omega_a$, $t(\Delta)$ is the single-atom amplitude transmission.
%Note that the spectrum of the transmitted light is independent from the position of the single atoms.
The model also allows us to calculate the probability amplitude of finding the n$^{\textrm{th}}$ atom in the excited state $\phi_{n}$: 
\begin{eqnarray}
\phi_{n}( \Delta )=i \dfrac{\sqrt{v_g}}{\sqrt{\beta \Gamma_0}}(t^{n}(\Delta)-t^{n-1}(\Delta)) ~,
\label{eq:ExcProb}
\end{eqnarray}
As mentioned in the main text, in the linear regime, the transmitted optical field in the time-domain after excitation with a pulse with scalar field amplitude $u_{in}(t)$ can be calculated as:
\begin{eqnarray}
u_{out}(t) = \mathcal{F}^{-1}[u_{in}(\Delta) t_N(\Delta)] ~,
\label{eq:Convolution}
\end{eqnarray}
where  $\mathcal{F}^{-1}$ indicates the inverse Fourier transform. A similar equation can be derived for the excitation amplitude for each of the $N$ atoms.

\subsection{Bidirectional model}
The general case of N atoms and  bidirectional coupling, in which photons are allowed to propagate both in the forward and backward direction, can be described with the following Hamiltonian:
\begin{eqnarray}
\dfrac{H}{\hbar}= \int_{-\infty}^{\infty} \bigg[ a_{x, \rightarrow}^{\dagger} (-i v_g \dfrac{\partial}{\partial x}) a_{x, \rightarrow} + 
a_{x, \leftarrow}^{\dagger} (+i v_g \dfrac{\partial}{\partial x}) a_{x, \leftarrow} +
\nonumber\\*
\sum_{n=1}^N \delta(x-x_n) [ \Omega_{a,n}  \sigma_n^+\sigma_n^- +
V_{n}(\sigma_n^+ a_{x, \rightarrow} + \sigma_n^- a_{x, \rightarrow}^{\dagger} + 
\nonumber\\*
\sigma_n^+ a_{x, \leftarrow} + \sigma_n^- a_{x, \leftarrow}^{\dagger})]
\bigg] dx ~,
\qquad \qquad \qquad
\label{eq:HamiltonianBidirectional}
\end{eqnarray}
where $\rightarrow$ and $\leftarrow$ refer to the forward and backward directions and, otherwise, the operators are defined as in Eq. \ref{eq:Hamiltonian}.
%the operators $a_{x, \rightarrow / \leftarrow }^{\dagger}$ and $a_{x, \rightarrow / \leftarrow}$ respectively create and annihilates a photon at position $x$ propagating forward or backward, $\sigma_n^+$ and $\sigma_n^-$ are the raising and lowering operators for the $n^{th}$ atom at position $x_n$, $N$ is the total number of atoms and $v_g$ is the group velocity of light.
In addition, now we have the following relations: $\Omega_{a,n} = [\omega_a-i(1 - \beta_{n} \Gamma_0)$ and $V_{n} = \sqrt{ v_g \beta_{n} \Gamma_0}$, where $\omega_a$ is the atomic resonance frequency and $\beta_{n}$ indicates the ratio of the spontaneous emission rate of the $n^\textrm{th}$ atom into the waveguide and the total single-atom emission rate $\Gamma_0$.

In the limit of a single excitation, the generic state can be written as: $\ket{\psi} = \int_{-\infty}^{\infty} [  \varphi_{\rightarrow}(x)  a_{x, \rightarrow}^{\dagger} + \varphi_{\leftarrow}(x)  a_{x, \leftarrow}^{\dagger} + \sum_{n=1}^N \delta(x-x_n) \phi_{n} \sigma_n^+ dx ] \ket{0}$, where $\varphi_{\rightarrow}(x)$ and $\varphi_{\leftarrow}(x)$ are the probability amplitudes of finding a forward and backward propagating photon at position $x$, respectively.

\paragraph{Solution for N=1}
For a single atom (i.e., $N=1$) the bidirectional model yields the following transmission and reflection coefficients (respectively $t$ and $r$):
\begin{eqnarray}
t(\Delta)=1-\dfrac{\beta \Gamma_0}{\frac{\Gamma_0}{2} + i \Delta} ~,
\\
r(\Delta)=-\dfrac{\beta \Gamma_0}{\frac{\Gamma_0}{2} + i \Delta} ~.
\label{eq:SingleAtomTranmissionReflection}
\end{eqnarray}

\paragraph{Solution for arbitrary N}
In the case of arbitrary $N$, it is convenient to define the following two quantities: $t_n=\varphi_{\rightarrow,n+1}/\varphi_{\rightarrow,n}$ and $s_n=\varphi_{\leftarrow,n}/\varphi_{\rightarrow,n}$, where  $\varphi_{\rightarrow / \leftarrow,n+1}$ indicates the field right after the $n^\textrm{th}$ atom. Then, solving the Schr\"odinger equation, we derive the recursion formulae:
\begin{eqnarray}
t_{n}(\Delta)=1-\dfrac{\beta_{n} \Gamma_0 + \sqrt{\beta_{n} \beta_{n}} s_{n+1} e^{-2ikx_n} \Gamma_0} {\frac{\Gamma_0}{2} + i \Delta + \sqrt{\beta_{n} \beta_{n}} s_{n+1} e^{-2ikx_n} \Gamma_0} ~, \qquad
\label{eq:RecursiveFormulas_t}
\\
s_{n}(\Delta)=(t_n-1)e^{i2kx_n}+s_{n+1}t_n ~, \qquad \qquad \qquad  
\label{eq:RecursiveFormulas_s}
\end{eqnarray}
where $k$ is the wavevector of light. Since we excite the atoms from a single direction only (i.e.,  $s_{N+1}=0$), these equations can be solved for $n=N$ and, then, recursively for the remaining atoms.
Finally, the ensemble amplitude transmission and reflection can be computed as:
\begin{eqnarray}
t_{N}(\Delta)= \prod_{j=1}^N t_j(\Delta) ~,
\label{eq:SolutionsBidirectionalModel_t}
\\
r_{N}(\Delta)= s_1(\Delta) ~.
\label{eq:SolutionsBidirectionalModel_r}
\end{eqnarray}

\subsection{Comparison between uni- and bidirectional model in the forward direction}

In the main manuscript we mentioned that because of collective enhanced forward scattering, the unidirectional model suffices to correctly describe the shape of the transmitted pulses. As an example, Fig. \ref{fig:ComparisonUnidirectionalBidirectional} compares the predictions of the unidirectional (eq. \ref{eq:Transmission}) and bidirectional (eq. \ref{eq:SolutionsBidirectionalModel_t}) models for OD = 19.3 and $\Delta$=17.3 $\Gamma_0$ (same parameters as in Fig. 2a of the main manuscript). No difference is apparent.
The calculation in the bidirectional case has been performed by assuming an ensemble of atoms randomly distributed along the nanofiber with coupling coefficients $\beta=0.55\%$ (average value in our experiment \cite{Johnson2019}). The results in Fig. \ref{fig:ComparisonUnidirectionalBidirectional} are obtained averaging over $10^4$ random configurations.

\subsection{Addition of a ring-resonator to the model}
The addition of a fiber ring-resonator to the setup can be included in our model by appropriately modifying the system Hamiltonian (see supplemental material of Ref. \cite{Johnson2019} for more details). Limiting the analysis to a unidirectional model, we can write the amplitude transmission of the system:
\begin{eqnarray}
t_{cav,N}(\Delta)=\dfrac{t_{rt}{t_N(\Delta)}e^{ikL}-t_{c}}{t_{rt}t_{c}{t_N(\Delta)}e^{ikL}-1} ~,
\label{eq:TransmissionCavity}
\end{eqnarray}
where $t_{rt}$ indicates the cavity roundtrip amplitude transmission, $t_{c}$ is the amplitude transmission of the fibre coupler used to launch light into the cavity,  $L$ is the optical length of a cavity and $t_N$ is the amplitude transmission through the ensemble of $N$ atoms as calculated earlier.
The cavity parameters needed for the theoretical prediction in Fig. 5(b-h) of the main manuscript have been independently estimated by measuring the pulse propagation in the absence of the atoms.
The complete time trace of the cavity ringdown measurement of Fig. 5(b-h) is shown in Fig. \ref{fig:CavityComplete}.

\section{Time evolution of the atomic excitation probability}

To support the results presented in Fig. 2(a) of the main manuscript,  we include in Fig. \ref{fig:TransmissionVsDetuning_ColormapSingleAtoms}(a) and (b) the calculated probability of finding each atom in the ensemble in the excited state as a function of time for our experimental parameters (OD = 19.3 and $\Delta$ = 17.3 $\Gamma_0$ as well as OD$_{tot}$=56 and $\Delta$ = 8.7 $\Gamma_0$). Both graphs clearly show the growing complexity of the individual Rabi oscillations for the atoms along the array.

At the start of the pulse, the atoms only see the laser light and all Rabi-oscillations start in phase. With increasing time, the light seen by the individual atoms is not only given by the incoming laser but by the sum of the laser field and the field emitted by all previous atoms along the array. As a consequence, amplitude and phase of the light arriving at the individual atoms become strongly dependent on their position along the array. This changes the relative phase of the Rabi-oscillations of the atomic populations along the ensemble. For a given position along the ensemble, this can result in Rabi-oscillations with a phase that is flipped with respect to the oscillations of the first atoms in the ensemble. In the experiment, we measure the light transmitted through the ensemble which is the sum of the incident probe pulse and the light emitted by all the atoms into the waveguide. Here, the position-dependent phase of the atomic Rabi-oscillations manifests itself as phase flips of the oscillation observed in the transmitted light. When increasing the number of atoms coupled to the waveguide, these phase flips occur at earlier times. For very large ensembles, these phase flips can even occur within the first Rabi-cycle which then gives rise to collective oscillatory signals, whose frequency exceeds the Rabi-frequency $\Omega \approx \Delta$ of the individual atoms, cf. Fig. 5(e) of the main manuscript (OD$_{tot}$=56) and Fig. \ref{fig:TransmissionVsDetuning_ColormapSingleAtoms} (b) ($N \approx 2500$).

\begin{figure}[]
	\includegraphics[width=1\linewidth]{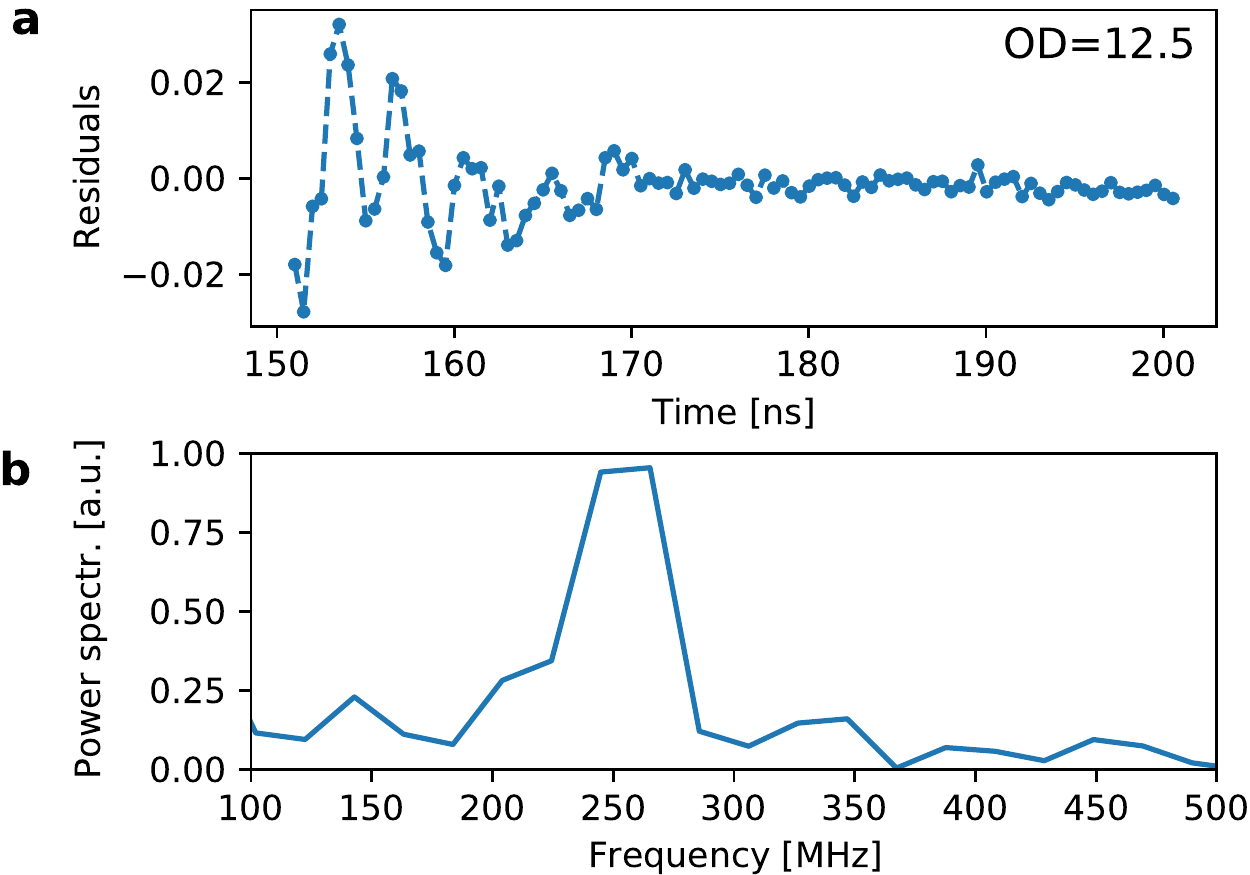}\
	\caption{\label{fig:Superradiance_ResidualsAndFT} (a) Residuals between the experimental data and an exponential fit for OD = 12.5 and (b) corresponding Fourier transform. A peak in the power spectral density is apparent at a frequency of $\approx$ 250 MHz.}
\end{figure}

%The atomic populations and thus the relative contribution of the different atoms to the transmitted light changes. As a consequence, from a given time on, the collectively emitted light is dominated by emission from atoms that undergo Rabi-oscillations with a large phase-difference with respect to the first atoms along the array.  With increasing atom number, this time at which the first phase flip occurs decreases and can even occur even before a single Rabi cycle is completed. This causes the fast oscillations in the transmitted power, whose frequency exceeds the detuning $\Delta$.
%With time, the position at which adjacent atoms oscillate with opposite phases moves against the direction of propagation of the light.
%It is interesting to compare these results with the calculation shown in Fig. \ref{fig:TransmissionVsDetuning_ColormapSingleAtoms}(b), in which we consider the parameters of Fig. 5(e) of the main manuscript (OD$_{tot}$=56 and $\Delta$ = 8.7 $\Gamma_0$). 
%The light detected in forward direction is the some of the initial laser pulse and the light emitted by the individual atoms. As can be seen 
%As a consequence, the light emitted by the parts of the ensemble that oscillate at opposite phase array will lead to a second maximum in the emitted light  

\section{Superradiance and quantum beats}

Figure \ref{fig:Superradiance_AllTheFits} shows the data, used to estimate the pulse decay rate as a function of the OD (see Fig. 3 of the main manuscript). The data up to OD=20.7 were fit over 30 ns, while for the following the fit range was reduced to 15 ns, because of their faster decay rate. 
Independently of the OD, small-amplitude oscillations are superimposed on the pulse decay. We attribute these clearly visible oscillations to quantum beats between the $6P_{3/2}$, $F^\prime = 4$ and $F^\prime = 5$ excited states. As an example, Fig. \ref{fig:Superradiance_ResidualsAndFT} shows the residuals between the experimental data and an exponential fit for OD = 12.5 as well as the corresponding Fourier transform. A clear peak in the power spectral density is apparent at a frequency of $\approx$ 250 MHz, which closely matches the frequency difference of 251.0 MHz between the above mentioned energy levels  \cite{Steck2019}.

\end{document}